\documentclass[draftcls,onecolumn,12pt]{IEEEtran}
\usepackage{graphicx,psfrag,epsfig,epsf,latexsym,hhline,amsmath,amssymb,multirow}
\usepackage{multirow}
\usepackage{epstopdf}
\usepackage{amsthm}
\usepackage{algorithm}
\usepackage{algpseudocode}
\usepackage{multicol}
\usepackage{cases}
\usepackage{blindtext}
\usepackage{etoolbox}
\usepackage{cite}
\usepackage{color}

\makeatletter
\patchcmd{\@makecaption}
  {\scshape}
  {}
  {}
  {}
\makeatother
\graphicspath{ {figures/} }
\theoremstyle{definition}

\algtext*{EndWhile}% Remove "end while" text
\algtext*{EndFor}% Remove "end while" text
\algtext*{EndIf}% Remove "end if" text
\newcommand{\defeq}{\triangleq}

\newcommand{\iid}{i.\@i.\@d.\ }

\theoremstyle{definition}\newtheorem{lemma}{Lemma}
\theoremstyle{definition}
\theoremstyle{definition}
\theoremstyle{definition}

\newtheorem{remark}[lemma]{Remark}

\newcommand\shortintertext[1]{%
	\ifvmode\else\\\@empty\fi
	\noalign{%
		\penalty0%
		\vbox{\mathstrut}%
		\penalty10000%
		\vskip-\baselineskip
		\penalty10000%
		\vbox to 0pt{%
			\normalbaselines
			\ifdim\linewidth=\columnwidth
			\else
			\parshape\@ne
			\@totalleftmargin\linewidth
			\fi
			\vss
			\noindent#1\par}%
		\penalty10000%
		\vskip-\baselineskip}%
	\penalty10000}

\begin{document}
\title{Joint Source-Channel Decoding of Polar Codes for Language-Based Source}
%\title{Joint list decoding of polar codes with language-based source}
\author{Ying Wang\dag, Minghai Qin\S, Krishna R. Narayanan\dag, Anxiao (Andrew) Jiang\ddag, and Zvonimir Bandic\S\\
\dag Department of Electrical and Computer Engineering, Texas A\&M University \\
\ddag Department of Computer Science and Engineering, Texas A\&M University\\ 
\S Storage Architecture, San Jose Research Center, HGST\\
{\tt\small {\{yingwang@tamu.edu, minghai.Qin@hgst.com,  krn@ece.tamu.edu,}}\\ {\tt\small {ajiang@cse.tamu.edu, and zvonimir.bandic@hgst.com\}} }

}

\maketitle

\begin{abstract}
We exploit the redundancy of the language-based source to help polar decoding. By judging the validity of decoded words in the decoded sequence with the help of a dictionary, the polar list decoder constantly detects erroneous paths after every few bits are decoded. This path-pruning technique based on joint decoding has advantages over stand-alone polar list decoding in that most decoding errors in early stages are corrected. In order to facilitate the joint decoding, we first propose a construction of dynamic dictionary using a trie and show an efficient way to trace the dictionary during decoding. Then we propose a joint decoding scheme of polar codes taking into account both information from the channel and the source. {The proposed scheme has the same decoding complexity as the list decoding of polar codes}. A list-size adaptive joint decoding is further implemented to largely reduce the decoding complexity. We conclude by simulation that the joint decoding schemes outperform stand-alone polar codes with CRC-aided successive cancellation list decoding by over 0.6~dB.

\end{abstract}

% Note that keywords are not normally used for peerreview papers.

\section{Introduction}
Shannon's theorem \cite{shannon1948} shows that separate optimization for source and channel codes achieves the global optimum. However, it is subject to impractical computational complexity and unlimited delay. In practice,  joint source and channel decoding (JSCD) is able to do much better than separate decoding if the complexity and delay constraints exist. It is based on the fact that there is still redundancy left in the source after compression. The idea is to exploit the source redundancy to help with channel decoding. In particular for language-based source, a lot of features can be exploited such as the meaning of words, grammar and syntax.

Great efforts have been put into joint decoding. Hagenauer \cite{hagenauer1995source} estimates residual correlation of the source and does joint decoding with soft output Viterbi algorithm. In \cite{buttigieg1995map}, soft information is used to perform the maximum \textit{a posteriori} decoding on a trellis constructed by the variable-length codes (VLCs). Joint decoding of Huffman codes and Turbo codes is proposed in \cite{guivarch2000joint}. A low-complexity chase-like decoding of VLCs is given in \cite{zribi2012low}. However, few works have been done for JSCD specifically for language-based source. In \cite{jiang2015enhanced},  LDPC codes are combined with a language decoder to do iterative decoding and it is shown to achieve a better performance.

Polar codes are gaining more attention due to the capacity achieving property \cite{arikan2009channel} and advantages such as low encoding/decoding complexity and good error floor performance \cite{eslami2013finite}.  However, successive cancellation (SC) decoding for finite length polar codes is not satisfying\cite{eslami2013finite}. To improve the  performance, belief propagation (BP) decoding of polar codes is proposed in \cite{hussami2009performance} with limited improvement over additive white Gaussian noise (AWGN) channels. It is further improved by a concatenation  and iterative decoding \cite{ywang2015concat}\cite{guoenhanced}. Successive cancellation list (SCL) decoding can provide substantial improvement over SC decoding.  It is shown in \cite{tal2011list} that with list decoding, the concatenation of polar codes with a few bits cyclic redundancy check (CRC) can outperform LDPC codes in WiMax standard. We denote the concatenated codes as polar-CRC codes and the decoding of such codes as CRC-aided SCL decoding.
 
\begin{figure}
  \centering
  \includegraphics[width=\columnwidth]{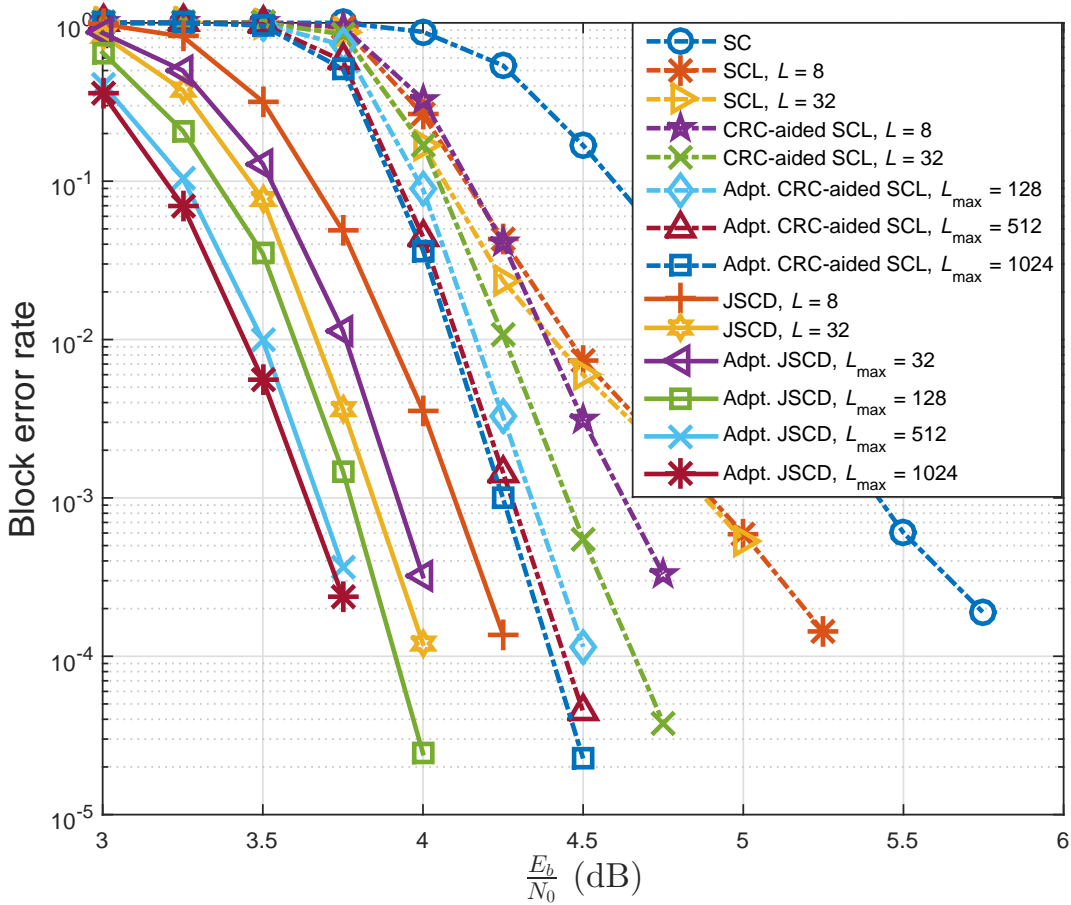}
  \caption{Block error rate of different decoding schemes over AWGN channels: a) SC decoding; b) SCL decoding ($L=8,32$); c) CRC-aided SCL decoding ($L=8,32$); d) Adapt. CRC-aided SCL decoding ($L_{\textrm{max}}=128,512,1024$); e) Joint source channel decoding ($L=8,32$); f) List-size adaptive JSCD ($L_{\textrm{max}}=32,128,512,1024$). All codes have $n =8192$ and $k =7561$.}
      \label{fig:pe_lp_awgn}
\end{figure}

\begin{figure}
	\centering
	\includegraphics[width=\columnwidth]{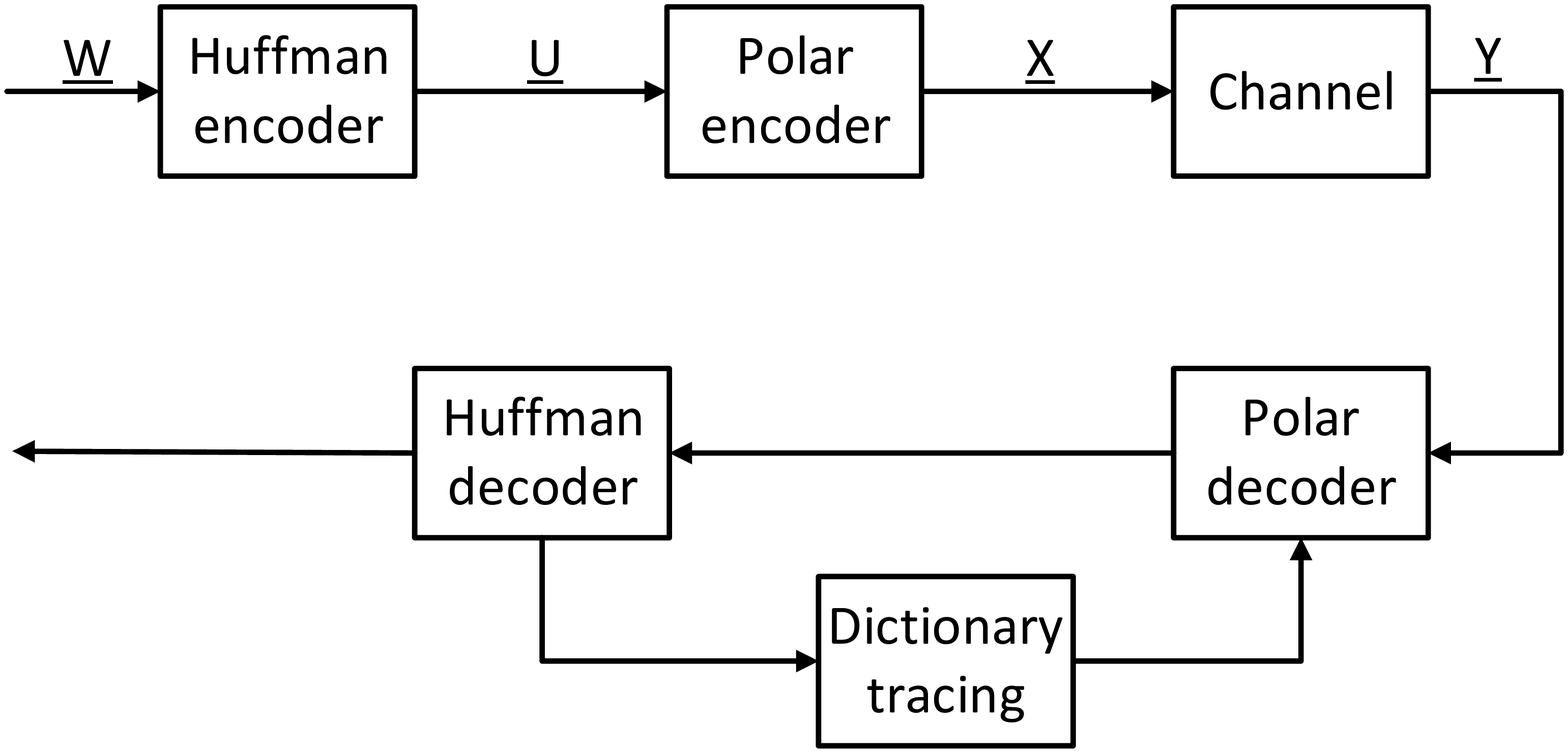}
	\caption{A system model for joint source-channel decoding}
	\label{fig:framework}
\end{figure}

In this paper, we manage to improve the decoding of polar codes by exploring the source redundancy. We propose a joint list decoding scheme of polar codes with \textit{a priori} information from a word dictionary. Fig.~\ref{fig:pe_lp_awgn} shows a block error rate comparison of different polar decoders for transmitting English text as sources. It is observed that over 0.6~dB gain can be achieved by joint source-channel decoding over stand-alone CRC-aided SCL decoding with comparable code parameters.   Fig.~\ref{fig:framework} illustrates the framework of the proposed coding scheme. We consider text in English, and the extension to other languages is straightforward. In our framework, the text is first compressed by Huffman codes and then encoded by polar codes. On the decoder side, the received sequence is jointly decoded by the polar code and a language decoder. The language decoder consists of Huffman decoding and dictionary tracing. It checks the validity of the decoded sequence by recognizing words in the dictionary. 
The language decoder has a similar function as CRCs for polar codes. Instead of picking a valid path from the list, the language decoder uses a word dictionary to select most probable paths, where the word dictionary can be viewed as local constraints on the decoded subsequences. A critical advantage of the language decoder over global CRC constraints is that it can detect the validity of partially decoded paths before decoding the whole codeword.  In this way, incorrect paths can be pruned at early stages, resulting in a larger probability that the correct path survives in the list. 

{The rest of the paper is organized as follows. In Section~\ref{sec:basics}, the basics of polar codes and the list decoder are reviewed.  In Section~\ref{sec:jscd}, the proposed joint source-channel decoding scheme for polar codes is presented.  Simulation results are presented in Section~\ref{sec:simulation}. In Section~\ref{sec:Lang_stat}, a brief discussion on the statistics of English language and advantages of JSCD is presented and we conclude the paper in Section~\ref{sec:conclusion}.}

\section{Backgrounds} \label{sec:basics}
In this section, we give a brief review of polar codes and two decoding algorithms, namely, SC decoding and SCL decoding. Throughout the paper, we will denote a vector $(x_i,x_{i+1},\ldots,x_j)$ by $x_i^j$, denote the set of integers $\{1,2,\ldots,n\}$ by $[n]$, denote the complement of a set $F$ by $F^c$, and denote probability measurement by $P(\cdot)$.

\subsection{Polar codes}
Polar codes are recursively encoded with the generator matrix $G_n=R_nG_2^{\otimes m}$, where $R_n$ is a $n\times n$ bit-reversal permutation matrix, $G_2=\left[ \begin{array}{cc}
1 & 0  \\ 
1 & 1   \end{array} \right]$, and $\otimes$ is the Kronecker product. The length of the code is $n=2^m$. Ar{\i}kan's channel polarization principle consists of two phases, namely channel combining and channel splitting. Let $u_1^n\defeq u_1 u_2\ldots u_n$ be the bits to be encoded, $x_1^n\defeq x_1 x_2 \ldots x_n$ be the coded bits, and $y_1^n\defeq y_1 y_2 \ldots y_n$ be the received sequence. Let $W(y|x)$ be the transition probability of a binary-input discrete memoryless channel (B-DMC). For channel combining, $N$ copies of the channel are combined to create the channel
\begin{equation*}
\label{eq:equivalentchannel}
W_n(y_1^n|u_1^n)\defeq W^n(y_1^n|u_1^nG_n) = \prod_{i=1}^n W(y_i|x_i),
\end{equation*}
where the last equality is due to the memoryless property of the channel. The channel splitting phase splits $W_n$ back into a set of $n$ bit channels
\begin{equation*}
\label{eq:ithequivalentchannel}
W_n^{(i)}(y_1^n,u_1^{i-1}|u_i)\defeq \frac{1}{2^{n-1}}\sum_{u_{i+1}^n} W_n(y_1^n|u_1^n),~i=1,\ldots,n.
\end{equation*}
Let $I(W)$ be the channel capacity of $W$. The bit channels $W_n^{(i)}$ will polarize in the sense that a fraction of bit channels will have $I(W_n^{(i)})$ converging to 1 as $n\rightarrow\infty$ and the other fraction will have $I(W_n^{(i)})$ converging to 0 as $n\rightarrow\infty$. Ar{\i}kan shows in \cite{arikan2009channel} that for the binary-input discrete memoryless channels, the fraction of $I(W_n^{(i)})$ converging to 1 will equal $I(W)$, the capacity of the original channel.

With channel polarization, the construction of Ar{\i}kan's polar codes is to find a set of bit channel indices  $F^c$ with highest quality and transmit information only through those channels. The remaining set of indices $F$ are called frozen set and the corresponding bits are set to fixed values known to the decoder.  It is proved in~\cite{arikan2009channel} that under SC decoding, polar codes asymptotically achieves the capacity of B-DMC channels. If the frozen bits are all set to 0, the SC decoder makes decisions as follows: $\hat{u}_i=0$ if $i \in F$; otherwise,
\begin{numcases}{\hat{u}_i=}\nonumber
	0, \;\text{if $L_n^{(i)}(y_1^n,\hat{u}_1^{i-1})\geq 0$}\\\nonumber
	1, \;\text{otherwise}	
\end{numcases}
where $L_n^{(i)}(y_1^n,\hat{u}_1^{i-1})$ is the log-likelihood ratio (LLR) of each bit $u_i$
\begin{equation}
    L_n^{(i)}(y_1^n,\hat{u}_1^{i-1})=\log \frac{W_n^{(i)}(y_1^n,\hat{u}_1^{i-1}|u_i=0)}{W_n^{(i)}(y_1^n,\hat{u}_1^{i-1}|u_i=1)}. \label{equ:LLR}
\end{equation}
Ar{\i}kan has shown that Eq.~(\ref{equ:LLR}) admits a recursive structure with decoding complexity $O(n\log n)$. The block error rate $P_B$ of polar codes
satisfies $P_B\leq 2^{-n^\beta}$ for any $\beta<\frac{1}{2}$ when the block length $n$ is large enough \cite{arikan2009rate}.

\subsection{List decoding of polar codes}
The SC decoder of polar codes makes hard decision of the bit in each stage. This may lead to severe error propagation problems. Instead, the SCL decoder keeps a list of the most probable paths. In each stage, the decoder extends the path with both 0 and 1 for unfrozen bit and the number of paths doubles. Assume the list size is $L$. When the number of paths exceeds $L$, the decoder picks $L$ most probable paths and prunes the rest. After decoding the last bit, the most probable path is picked as the decoded path. The complexity of SCL decoding is $O(Ln\log n)$, where $n$ is the block length of the code. An extra improvement can be brought by SCL decoding with CRC, which increases the minimum distance of polar codes and helps to select the most probable path in the list. The adaptive SCL decoder with a large list size can be used to fully exploit the benefit of CRC while largely reducing the decoder complexity \cite{li2012adaptive}.

\section{Joint source channel decoding}\label{sec:jscd}
In this section, we provide a detailed description of the proposed joint source channel coding scheme. We will first illustrate the decoding rule mathematically and then explain the derivation of each term in the equations algorithmically. 

The maximum \textit{a posteriori} (MAP) decoding aims to find $\max_{u_1^n} P(u_1^n|y_1^n)$. To avoid exponential complexity in $n$, SCL decoding tries to maximize $P(u_1^i|y_1^n),i = 1,\ldots,n$ progressively by breadth-first searching a path in the decoding tree, where for each length-$i$ path, a constant number, often denoted by $L$, of most probable paths are kept to search for length-$(i+1)$ paths. Consider that 
\[
P(u_1^i|y_1^n) = \frac{P(u_1^i,y_1^n)}{P(y_1^n)} \propto P(y_1^n|u_1^i)P(u_1^i).
\]
By source-channel separation theorem, a stand-alone polar decoder calculates the first term $P(y_1^n|u_1^i)\propto P(y_1^n,u_1^{i-1}|u_i)$ by a recursive structure, assuming $u_1^i$ are independently and identically distributed (i.i.d.) Bernoulli$(0.5)$ random variables, and thus the second term can be obliterated since $P(u_1^i)=2^{-i},\forall u_1^i\in\{0,1\}^i$. However, in the language-based JSCD framework, $u_1^i$ are no longer i.i.d., one obvious correlation of which is that $u_1^i$ is feasible only if the decoded text, translated from $u_1^i$ by Huffman decoder, consists words in the dictionary. Therefore, $P(u_1^i)$ contributes critically in the path metric $P(u_1^i|y_1^n)$, and in particular, if $P(u_1^i)=0$, this path should be pruned despite the metric $P(y_1^n|u_1^i)$ obtained from the channel. This pruning technique enables early detection of decoding errors and is critical in keeping the correct path in the list. Algorithm~\ref{alg:jscd} shows a high-level description of JSCD. 

\begin{algorithm} [htbp]
	\caption{A high-level description of JSCD}
	\label{alg:jscd}
	\textbf{Input:} $y_1^n$, $L$
	
	\textbf{Output:} ${u}_1^n$
	\begin{algorithmic}[1]
		\State Initialize: $i\leftarrow 1$;  $l_\textrm{act}\leftarrow 1$; 
		\While{$i\leq n$}
		\If{$i\in F$}
		\State{$ u_i\leftarrow 0$} for each active path;
		\Else
		\State {$k \leftarrow 1$;}
		\For{each active path $l_j,j\in[l_\textrm{act}]$ }
		\For{$u_i = 0, 1$}
		\State Compute $P(y_1^n,u_1^{i-1}|u_i)$;
		\State Update $P(u_1^i)$;  \label{line:prior}
		\State $M_{k} \leftarrow  P(y_1^n,u_1^{i-1}|u_i)P(u_1^i)$ ;
		\State $k \leftarrow k+1$;
		\EndFor 
		\EndFor
		\State {$\rho \leftarrow \min(2l_\textrm{act}, L)$	;
		\State Keep most probable $\rho $ paths according to $M_{1}^{2l_\textrm{act}}$;
		\State $l_\textrm{act} \leftarrow \rho$;}
		\EndIf
		\State $i\leftarrow i+1$;					
		\EndWhile		
		\State Select the most probable path and output corresponding ${u}_1^n$.
	\end{algorithmic}
\end{algorithm}

Note that the only difference of Algorithm~\ref{alg:jscd} from the  SCL decoder of stand-alone polar codes presented in~\cite{tal2011list} is the introduction of $P(u_1^i)$ in~\ref{line:prior}th line. {Therefore}, the complexity of Algorithm~\ref{alg:jscd} is $O(Ln(\log n + C))$, where $C$ is the complexity of updating $P(u_1^i)$ from $P(u_1^{i-1})$. It will be shown later in Algorithm~\ref{alg:calculate_prior} that $C$ is a constant in $n$, i.e., the proposed JSCD algorithm has the same complexity as SCL decoders. The rest of this section is devoted to the data structures and algorithms to calculate $P(u_1^i),i=1,\ldots,n$.

Let $\cal A$ be the alphabet of symbols in text (e.g., $\{a,b,\ldots,z\}$ for lowercase English letters, $\{0,\ldots,127\}$ for symbols in ASCII table). {Let $\cal D$ be the set of words in the dictionary.} Since Huffman codes are instantaneously decodable,  we can represent $u_1^i$ in the concatenated form of $(w_1w_2\ldots w_{j-1}l_1 l_2 \ldots l_kr)$, where $w_1^{j-1}$ are $j-1$ uniquely decoded words in $\cal D$,  $l_1^k$ are $k$ uniquely Huffman-decoded symbols in $\cal A$ and $r$ is the remaining bit sequence.  Thus, we can represent $P(u_1^i)$ as follows
\begin{align}
P(u_1^i) &= P(w_1^{j-1} l_1^kr) \nonumber  \stackrel{\textcircled {\footnotesize 1}}{=} \prod_{m=1}^{j-1}P(w_m) P(l_1^k r) \nonumber\\
& = \prod_{m=1}^{j-1}P(w_m) \sum_{w} P(w), \label{eq:prior}
\end{align}
where in the summation, $w\in\cal D$ satisfies that in binary Huffman-coded representation, the first $k$ symbols equals $l_1^k$ and $r$ is a prefix of the remaining bit sequences. 
\begin{remark}
Note that the equality $\textcircled {\footnotesize 1}$ is under the assumption that words in a text are independent. This assumption is a first order approximation and more detailed study on Markov property of words in languages can be found in \cite{brown1992class}.
\end{remark}
\begin{remark}
The calculation of $P(u_1^i)$ should also take into account the probability of spaces (or punctuations) between words. This concern can be handled in two ways. One is to treat the space (or punctuations) as a separate word in the dictionary and estimate the probability of its appearance, the other is to treat the space (or punctuations) as a suffix symbol to all words in the dictionary. Although two approaches will result in different {values} of $P(u_1^i)$, the overall joint SCL decoding performance is similar. In our proposed algorithm, we use the latter solution. To simplify presentation of algorithms, we only append a space mark to all words.
\end{remark}
 
Now we focus on the efficient calculation of Eq.~(\ref{eq:prior}). Two trees are used to facilitate the calculation, one is a tree for Huffman coding and the other is a prefix tree (i.e., a trie) for tracing a partially decoded word in the dictionary.

\subsection{Trie representation of the dictionary}
\begin{figure}[htbp]
	\centering
	\includegraphics[width=\columnwidth]{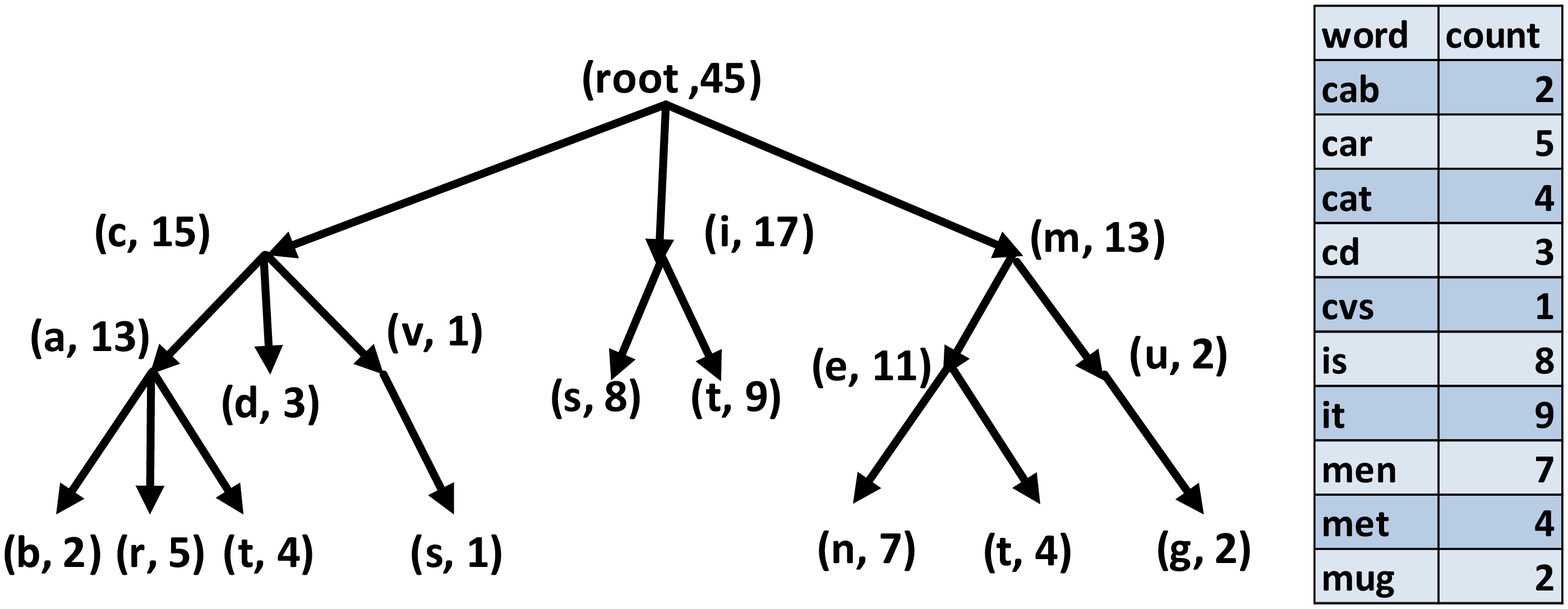}
	\caption{An illustrative example of a trie to represent the dictionary}
	\label{fig:dict_tree}
\end{figure}

A trie is an ordered tree data structure that is used to store a dynamic set or associative array where the keys are usually strings \cite{fredkin1960trie}. In our implementation, each node in the trie is instantiated as an object of a class named \texttt{DictNode}. As shown in Table~\ref{tab:dict}, it has 4 data members,  a symbol \texttt{c} (e.g., English letter), a variable \texttt{count} representing the frequency of the presence of this prefix, an indicator \texttt{is\_a\_word} indicating if the path from root to this node is a whole word\footnote{This data member can be omitted if spaces are appended to all words, but we keep it to present algorithms more clearly.}, and a vector of pointers \texttt{child[]} pointing to their children.   Fig.~\ref{fig:dict_tree} is an illustrative example of the dictionary represented by a trie. In an established trie, if the pointer that points to the end of a word (or a partial word) $w$ is known, then the calculation of $P(w)$  can be accomplished in $O(1)$ by dividing the count of the end node of the path associated with $w$ by the count of the root node.

\begin{table}[!htb]
    \begin{minipage}{.5\linewidth}
    \centering
      \caption{Data members of \texttt{DictNode} in $\cal T$}
  {\begin{tabular}{|r|r|}
\hline

    member &       type \\
\hline
\hline
         \texttt{c} &       \texttt{char} \\
\hline
     \texttt{count} &        \texttt{int} \\
\hline
 \texttt{is\_a\_word} &       \texttt{bool} \\
\hline
     \texttt{child[]} &  \texttt{DictNode*} \\
\hline
\end{tabular}} 
\label{tab:dict}
    \end{minipage}%
    \begin{minipage}{.5\linewidth}
      \centering
        \caption{Data members of \texttt{{H}uffNode} in $\cal H$}
  {
\begin{tabular}{|r|r|}
\hline
    member &       type \\
\hline
\hline
     \texttt{p} &        \texttt{double} \\
\hline
 \texttt{leftChild} &  \texttt{huffNode*} \\
\hline
\texttt{rightChild} &  \texttt{huffNode*} \\
\hline
    \texttt{symSet} &      \texttt{char*} \\
\hline
\end{tabular}} 
\label{tab:huff}
    \end{minipage} 
\end{table}

In order to establish the trie from extracted text (e.g., from books, websites, etc.), an algorithm with an inductive process can be used. That is, suppose we have a trie $\cal T$ that represents the first $i$ words of the extracted text, for the $(i+1)$st word $w=(l_1\ldots l_k)$ {(assuming it contains $k$ symbols)}, a pointer \texttt{p\_dict} is created to point to the root and the first symbol $l_1$ in the word is compared with the children of the root in $\cal T$. If $l_1$ exists as the symbol of a depth-1 node $m_1$, then \texttt{p\_dict} moves to $m_1$ and $l_2$ is compared with the children of $m_1$. The same operation continues until some $l_j$ does not exist in the children set of the node $m_{j-1}$ corresponding to path $l_1^{j-1}$. Then a new child with symbol $l_j$ is added to $m_{j-1}$ and the rest of the word $l_{j+1}^k$ is added accordingly. During the scan of $(i+1)$st word, the counts for
each node \texttt{p\_dict} visits are increased by 1.  Algorithm~\ref{alg:build_dict} shows the detail of the algorithm.

\begin{algorithm}
	\caption{Establish a trie for the dictionary from extracted text}
	\label{alg:build_dict}
	\textbf{Input:} a sequence of words $(w_1 w_2 \ldots w_N)$, each word is represented as a string\\
	\textbf{Output:} a trie $\cal T$
\begin{algorithmic}[1]
\State Initialize: Create a root node of $\cal T$ as an object of \texttt{DictNode};	
\For{$k=1$ to $N$}
		\State Let \texttt{p\_dict} point to the root of $\cal T$;
	\For{$i=1$ to the length of $w_k$}
			\If{\texttt{*p\_dict} has no child or $w_k[i]$ is not in the children set of \texttt{*p\_dict}}
			\State Create a new node as an object of \texttt{DictNode} with \texttt{c $\leftarrow$} $w_k[i]$, \texttt{count $\leftarrow$ 1} and \texttt{is\_a\_word $\leftarrow$ False};
			\State Insert the new node as a child of \texttt{*p\_dict};
			\State Move \texttt{p\_dict} to the new node;
			\If{$i==$ the length of $w_k$}
				\State \texttt{p\_dict->is\_a\_word $\leftarrow$ True};
				\EndIf
			\Else
			\State Find $j$, s.t. $w_k[i]==$\texttt{p\_dict->child[j]->c};
				\State \texttt{p\_dict->count++};
				\State \texttt{p\_dict $\leftarrow$ p\_dict->child[j]};
			\EndIf
	\EndFor		
%			\State Create a new \texttt{DictNode} with \texttt{c=space};
%			\State Insert the new node as a child of \texttt{*p\_dict};
%			\State \texttt{*p\_dict.Is\_a\_word=True};
\EndFor
\end{algorithmic}
\end{algorithm}

Since searching for a symbol as a child of a node in $\cal T$ can be accomplished in $O(1)$ using Hash table (e.g., \texttt{unordered\_map} STL container in C++), the time complexity of establishing the trie would be $O(N_\textrm{length}N_\textrm{word})$, where $N_\textrm{length}$ is the average length of a word and $N_\textrm{word}$ is the number of words extracted from some resource.

\subsection{Tree representation of Huffman codes }
\begin{figure}[htbp]
	\centering
	\includegraphics[width=\columnwidth]{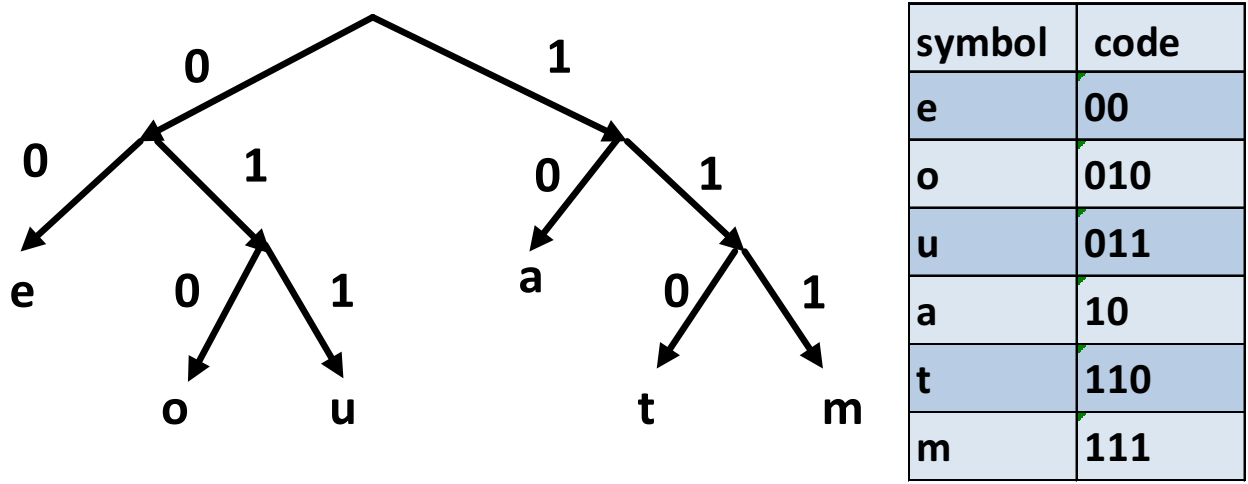}
	\caption{An illustrative example of a Huffman tree.}
	\label{fig:huff_tree}
\end{figure}

The Huffman codes for source coding  are for 1-grams, namely characters, or more specifically, letters and space mark. In principle, we can also build a Huffman code for $n$-grams. The Huffman codes are represented as a binary tree. Each node in the tree is instantiated as an object of a class \texttt{HuffNode} whose members are shown in Table~\ref{tab:huff}. 
In a typical Huffman tree realization, a node $m$  consists of three members: the probability \texttt{p} of the associated symbol and two pointers to their left and right children (\texttt{leftChild} and \texttt{rightChild}). In addition, we implement a fourth data member \texttt{symSet}, that is, a set of symbols that are descendants of $m$.  This extra data member helps in simplifying the calculation of Eq.~(\ref{eq:prior}) in the following manner. Note that in Eq.~(\ref{eq:prior}), $P(l_1^k r)$, the probability of a partial word is required. Assume $l_1^k$ is a path that ends in a node $n_k$ in the trie-represented dictionary $\cal T$  and $r$ is a path that ends in a node $n_r$ in the Huffman tree $\cal H$. Then $P(l_1^k r)$ can be calculated by summing up the counts (or probability) of the subset of children of $n_k \in \cal T$, such that the symbols associated with this subset are all descendants of $n_r\in\cal H$. By associating all descendants of $n_r$ as a data member to the node itself, the complexity of calculating $P(l_1^k r)$ is linear in the number of descendants of $n_r$, which is typically a small number and decreases exponentially in the depth of $n_r$. Fig.~\ref{fig:huff_tree} shows an illustrative example of a Huffman tree.

\subsection{Calculation of $P(u_1^i)$ with $\cal T$ and $\cal H$}
Next, we will present an algorithm to calculate $P(u_1^i)$ progressively according to Eq.~(\ref{eq:prior}). In each of $\cal T$ and $\cal H$, two pointers, denoted by \texttt{p\_dict} and \texttt{p\_huff}, are used respectively to locate the current decoding stages $i\in[n]$. They are initiated to point to the root of $\cal T$ and $\cal H$, respectively. A simple description of the algorithm is as follows. Let $u_1^{i-1}$ be represented as $(w_1^{j-1} l_1^k r)$ and suppose each term in  Eq.~(\ref{eq:prior}) is known up to index $i-1$. Suppose \texttt{p\_dict} and \texttt{p\_huff} point to two nodes in $\cal T$ and $\cal H$. To update $P(u_1^{i})$, first, \texttt{p\_huff} moves to its left or right child according to $u_i$. Let $\cal S$ be all descendant symbols of \texttt{*p\_huff}. Replace $P(l_1^k r)$ by the summation of probabilities associated with a set of children, denoted by $\cal C$, of \texttt{*p\_dict} such that $\forall  a \in \cal C$, the symbols associated with $a$ belongs to $\cal S$; If \texttt{*p\_huff} is a leaf, then \texttt{p\_dict} moves to its child according to the symbol \texttt{*p\_huff} associates and \texttt{p\_huff} is reset to point to the root of $\cal H$. If the symbol \texttt{*p\_huff} associates to does not exist in the children of \texttt{*p\_dict}, that means  $P(u_1^{i})$ should be set to 0 and this path has a decoding error and thus be pruned.  If furthermore \texttt{*p\_dict} is an end node of a word in $\cal T$, replace $P(l_1^k r)$ by $P(w_j)$ and \texttt{p\_dict} is reset to point to the root of $\cal T$. Let the multiplication of probabilities in Eq.~(\ref{eq:prior}) be denoted by $P_{wd}$, i.e., $P_{wd}=\prod_{m=1}^{j-1} P(w_m)$, where $P_{wd}$ can be  updated recursively.  A detailed description of this algorithm is presented in Algorithm~\ref{alg:calculate_prior}.

\begin{algorithm}
	\caption{Update $P(u_1^{i})$}
	\label{alg:calculate_prior}
	\textbf{Input:} $u_{i}$, $\cal T$, $\cal H$, \texttt{p\_dict}, \texttt{p\_huff}, $P_{wd}$ \\
	\textbf{Output:} \texttt{p\_dict}, \texttt{p\_huff}, $P(u_1^{i})$, $P_{wd}$
	\begin{algorithmic}[1] 	
		\State $\cal S \leftarrow $\texttt{TraceHuffmanTree($\cal H$,p\_huff,$u_{i}$)};
		\State $\cal C \leftarrow $\texttt{TraceDict($\cal T$,p\_dict,$\cal S$)};
		\State $P(l_1^k r) \leftarrow \sum_{w\in \cal C} P(w)$;
		\State $P(u_1^{i}) \leftarrow P_{wd}\cdot P(l_1^k r)$;
		\If{\texttt{p\_huff} points to a leaf in $\cal H$}		
		\State Move \texttt{p\_dict} to its child according to \texttt{p\_huff} ;
		\State Move \texttt{p\_huff} to the root of $\cal H$;
		\If{\texttt{p\_dict} points to a leaf in $\cal T$}
		\State $P(w_j) \leftarrow P(l_1^k r)$;
		\State $P_{wd} \leftarrow P_{wd}\cdot P(w_j)$;
		\EndIf
		\EndIf
	\end{algorithmic}
\end{algorithm}

\begin{algorithm}
	\caption{\texttt{TraceHuffmanTree($\cal H$,p\_huff,$u_{i}$)}}
	\label{alg:trace_huff_tree}
	\textbf{Input:} $u_{i}$, $\cal H$,  \texttt{p\_huff} \\
	\textbf{Output:} $\cal S$
	\begin{algorithmic}[1] 	
		\If{$u_{i}==0$}
			\State Move \texttt{p\_huff} to its left child;
		\Else
		\State Move \texttt{p\_huff} to its right child;
		\EndIf
			\State $\cal S \leftarrow$ \texttt{p\_huff}\texttt{->symSet};
	\end{algorithmic}
\end{algorithm}

\begin{algorithm}
	\caption{\texttt{TraceDict($\cal T$,p\_dict,$\cal S$)}}	\label{alg:trace_dict_tree}
	\textbf{Input:} $\cal T$, \texttt{p\_dict}, $\cal S$ \\
	\textbf{Output:} $\cal C$
	\begin{algorithmic}[1]
	\State $\cal C \leftarrow \emptyset$; 	
		\For{each symbol $s\in \cal S$}
		\If{$s$ is found in the children set of \texttt{*p\_dict}}
		\State $s$ is added to $\cal C$;
		\EndIf
		\EndFor
	\end{algorithmic}
\end{algorithm}

The complexity of Algorithm~\ref{alg:calculate_prior} involves operations of the two pointers. It takes $O(1)$ to extract the descendants of \texttt{*p\_huff} and it takes at most $O(N_\textrm{child})$ to sum up their probabilities, where $N_\textrm{child}$ is the number of children of a node in $\cal T$. Therefore, updating $P(u_1^i)$ is constant in $n$.

\subsection{List-size adaptive JSCD}
In order to simplify the SCL for JSCD, we implement the list-size adaptive SCL decoders as in~\cite{li2012adaptive}.  A few CRC bits are added for error detection. The adaptive SCL decoders start with $L =1$ and end up with an estimate $u_1^n$. If $u_1^n$ satisfies the CRCs, then $u_1^n$ are output as the decoded bits, otherwise, the list size doubles and the SCL JSCD is repeated. This process continues until $u_1^n$ satisfies the CRCs for some $L_\textrm{success}$ or the list size reaches a threshold $L_\textrm{max}$.

\section{Numerical results}\label{sec:simulation}
In this section, we present some numerical results that show the superiority of SCL JSCD over the stand-alone SCL decoder.

%\subsection{Dictionary}
%The dictionary is built from about 10 million extracted  words in Wikipedia pages. According to our simple analysis, the top 3000 most frequent words take $81\%$ of the probability.

\subsection{Dictionary}
The dictionary is built from about 10 million extracted  words in Wikipedia pages. {According to a word frequecy analysis in \cite{freqwords}, the top 3000 most frequent words take $81\%$ of the probability. }

\subsection{Polar codes and channel parameters}
In our simulation, the length of polar codes is fixed to be $n = 8192$ and the rate of the code is $0.923$ with data bits $k = 7561$. Two typical B-DMCs are assumed, namely, AWGN channels and binary symmetric channels (BSCs). The polar code used for AWGN channels is constructed by {density evolution in~\cite{mori2009performance}} at $\frac{E_b}{N_0}=4$ dB. The polar code used for BSCs is similarly constructed for a BSC with cross-over probability $0.003$. Six decoders of polar codes are compared for AWGNs, including a) successive cancellation decoders, b) stand-alone SCL decoders, c) stand-alone SCL decoders with 16-bit CRCs, d) adaptive CRC-aided SCL decoders, e) SCL decoders using JSCD, and f) list-size adaptive SCL decoders using JSCD. A subset of these decoders are compared for BSCs.

\subsection{Results}
\begin{figure}[htbp]
	\centering
	\includegraphics[width=\columnwidth]{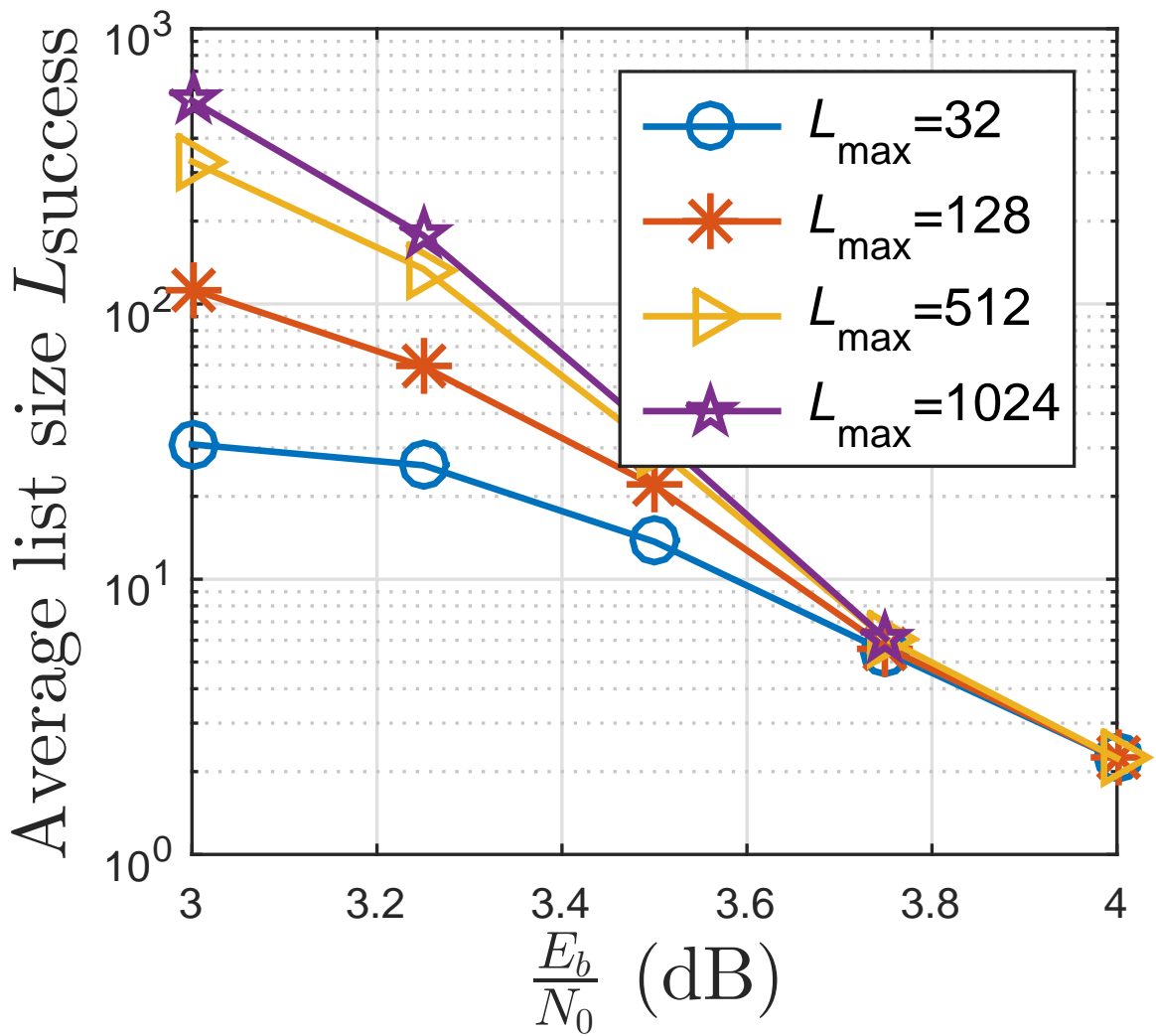}
	\caption{Average list size of adaptive JSCD}
	\label{fig:avg_list}
\end{figure}

Fig.~\ref{fig:pe_lp_awgn} shows a comparison of different decoders for AWGN channels. It can be seen that at block error rate of below $10^{-3}$, more than 0.6~dB gain over stand-alone CRC-aided SCL decoders can be realized by the list-size adaptive SCL JSCD decoders.  It is observed in our simulation that $L=1024$ would be large enough such that further increase of the list size will not contribute much to the performance. The decoding complexity of the list-size adaptive SCL JSCD is much lower than the complexity of SCL JSCD with fixed list size. Fig.~\ref{fig:avg_list} shows that the average list size $L_\textrm{success}$ decreases dramatically with the increase of SNRs. It is observed that at $\frac{E_b}{N_0}=4$ dB,  $L_\textrm{success}=2.24$ for all $L_\textrm{max}=128,512,1024$.

\begin{figure}[htbp]
	\centering
	\includegraphics[width=\columnwidth]{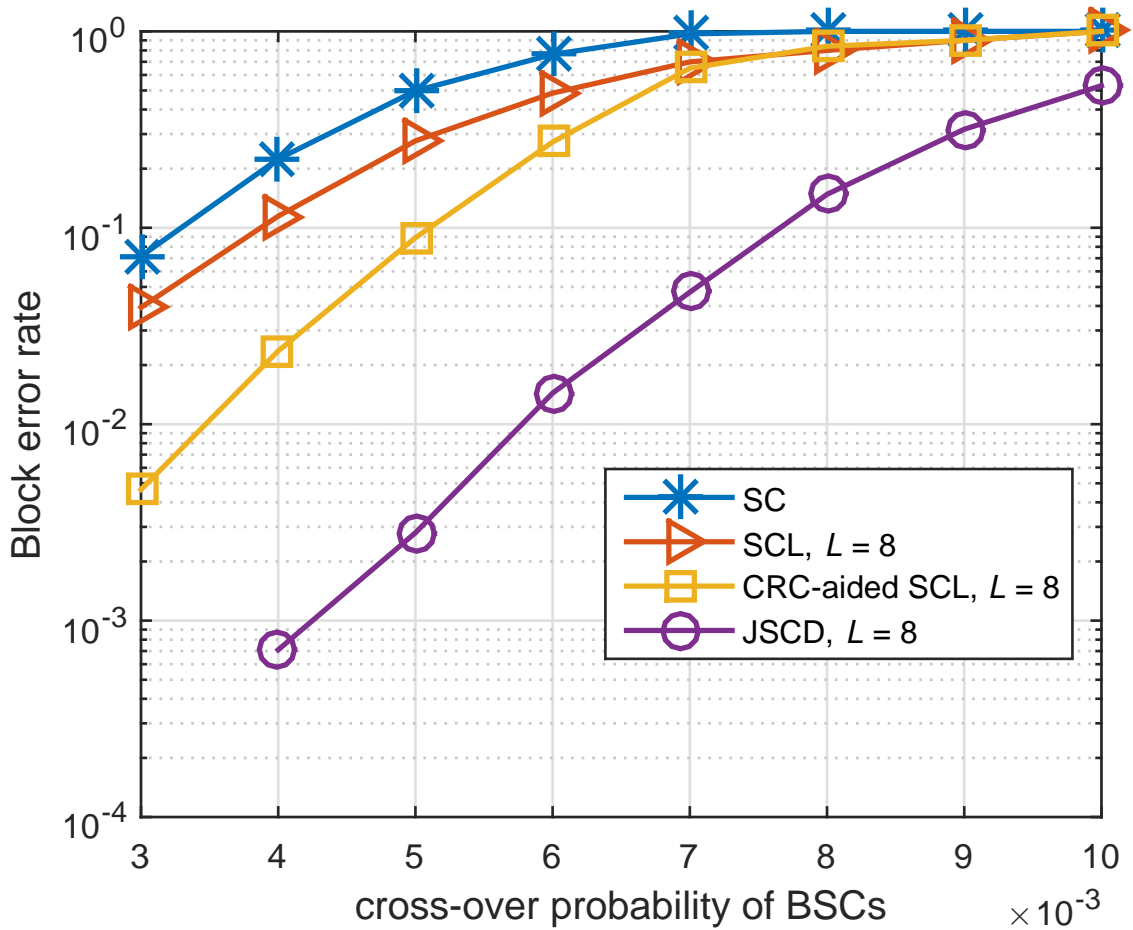}
	\caption{Comparison of different decoding schemes over BSCs}
	\label{fig:pe_lp_bsc}
\end{figure}

In order to show that the improvement of the proposed JSCD algorithm is not channel specific, Fig.~\ref{fig:pe_lp_bsc} shows a comparison of 4 decoders for BSCs. {The results consistently show the superiority of JSCD scheme over CRC-aided SCL decoding. }

\section{Discussion on language statistics}\label{sec:Lang_stat}
In this section, some properties of language-based source is discussed to explain the significant gain achieved by JSCD with a dictionary. We can also identify the redundancy of the source and sparsity of words based on those properties.

\subsection{Redundancy of Huffman-encoded text}
%The optimal compression scheme would make the average bit length of the symbol equal the entropy of the source. That is
%\begin{align*}
%E[L]=H(S)
%\end{align*}
%\begin{align*}
%\text{H(X)}=-\sum_i p_i\log_2 p_i
%\end{align*}
%where $p_i$ is the probability of the $i$-th symbol.
The language has diverse features including semantics, grammar and syntax, \textit{etc}. {From the fundamental coding theorem, the average number of bits to represent a word is lower bounded by the entropy of words.}
Shannon estimated the word entropy of printed English to be 11.82 bits per word \cite{shannon1951prediction}. Practically, we have collected a large number of words from extracted text and computed the entropy of words by
\begin{align*}
H(X)=-\sum_i p_i\log_2 p_i,
\end{align*}
where $H(X)$ is the entropy of the source $X$ and $p_i$ is the probability of the $i$th unique word, assuming that the words are independent. In our extracted text, the resultant entropy of words is estimated to be 10.41 bits per words. However, the average number of bits for a Huffman-encoded word is approximately 37 bits per words, which is much larger than both estimates, showing great redundancy remains in the compressed text. The major reason for such redundancy is that Huffman codebook is generated by the distribution of English letters instead of words, where strong correlation between letters exists. Some other factors to cause the difference in the length of Huffman-encoded word and the entropy of words include integer-length constraint of Huffman codes and mismatch between the source model and the actual text transmitted.

\subsection{Sparsity of words}
\begin{figure}[htbp]
	\centering
	\includegraphics[width=\columnwidth]{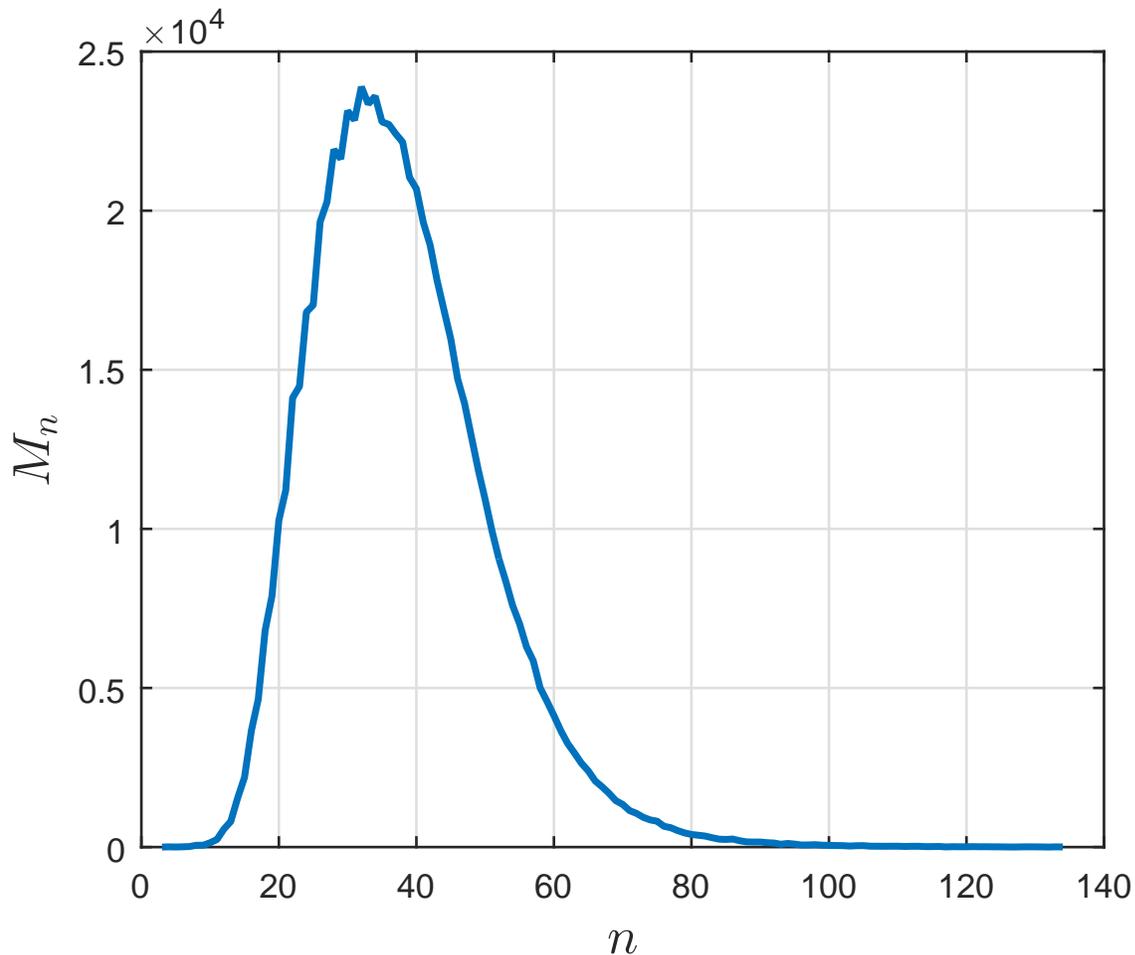}
	\caption{The number of words with length-$n$ Huffman-encoded binary sequences}
	\label{fig:number_words}
\end{figure}

\begin{figure}[htbp]
	\centering
	\includegraphics[width=\columnwidth]{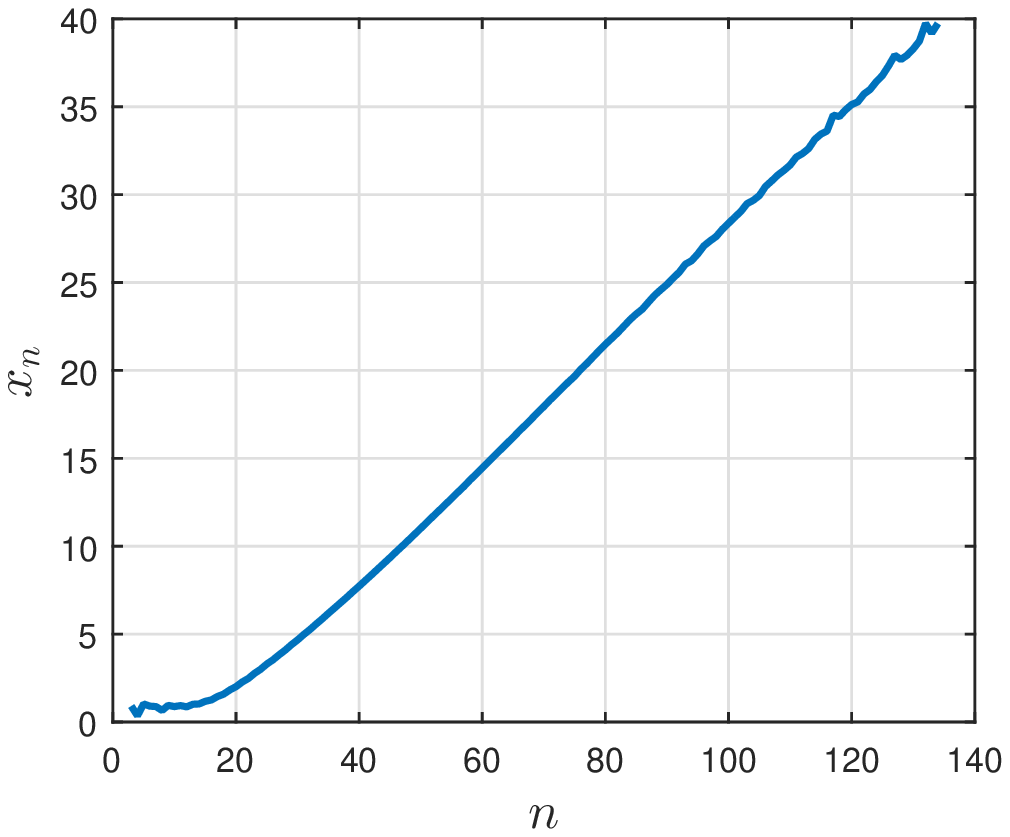}
	\caption{The sparsity exponent $x_n$, where $P_n=\frac{M_n}{2^n}=10^{-x_n}$ is the probability that a uniformly and randomly chosen binary sequence of length $n$ corresponds to a valid word. }
	\label{fig:de_words}
\end{figure}

Let $M_n$ denote the number of Huffman-encoded binary sequences of length $n$ that correspond to a word in the dictionary. We call such a sequence a valid binary sequence. Let $P_n$ be defined as
\[
P_n=\frac{M_n}{2^n},
\]
i.e.,  $P_n$ is the probability that a uniformly and randomly chosen binary sequence of length $n$ corresponds to a valid word. We can write $P_n$ in an exponential form
\[
P_n = 10^{-x_n},
\]
where $x_n$ represents the growth rate of sparsity of valid binary sequences. Based on statistics of the extracted text, $M_n$ and $x_n$ are shown in  Fig.~\ref{fig:number_words} and Fig.~\ref{fig:de_words}, respectively. Fig.~\ref{fig:number_words} illustrates that the  length of Huffman-encoded binary sequence for more than 97\% of words is concentrated between 15 and 70. Fig.~\ref{fig:de_words} shows that $x_n$ increases almost linearly in $n$ for $n>15$, thus $P_n$ decreases exponentially in $n$. Therefore, if $n$ is large, $P_n$ is very small, meaning valid binary sequences are sparse. The sparsity of valid words indicates that once the decoded binary sequence corresponds to a valid word, there is a high probability that the decoding is correct\footnote{In fact, for valid words with length-$n$ Huffman-encoded sequences, the Hamming distance and the probability of the words determine the error probability. The sparsity of valid words is a necessary condition for a large Hamming distance.  }.

\subsection{JSCD advantages over stand-alone polar decoding}
Based on the discussion above, we summarize the advantages of JSCD as follows. The redundancy left in the Huffman-encoded words can be exploited by the joint decoding of information provided by the source and the channel. This is achieved by the feedback of dictionary tracing in Fig.~\ref{fig:framework}, where words in the dictionary can be viewed as local constraints for subsequences in polar coded binary sequences. The sparsity of words suggests that the local constraints are efficient in pruning incorrect decoded subsequences. Opposed to global CRCs that select a correct path after the whole binary sequence is decoded, JSCD can select and prune the paths in early stages, resulting in a larger probability that the correct path survives in the list.

\section{Conclusion}\label{sec:conclusion}
In this paper, we exploit the redundancy in the language-based source to help polar decoding. We propose a joint decoding
scheme of polar codes taking into account the source information using a dictionary. A dynamic dictionary is constructed using a trie, and an efficient way to trace
the dictionary during decoding is proposed. {The decoding complexity is the same as list decoding of stand-alone polar codes.} Simulation results show that our scheme significantly outperforms  list decoding of CRC-aided polar codes. Further improvement is achieved by list-size adaptive joint decoding, while the decoding complexity is largely reduced.   The results indicate high efficiency of source redundancy in error detection and path pruning.

\bibliographystyle{IEEEtran}
\bibliography{jscd_full_v1.3}
\end{document}